\documentclass[usenatbib]{mn2e}
\usepackage{epsfig}
\usepackage{rotating}
\usepackage{multirow}
\usepackage{longtable}
\usepackage{breqn}

\bibpunct[, ]{(}{)}{;}{a}{}{,}

\title[Star-Formation in the ULIRG F00183-7111]{Star-Formation in the Ultraluminous Infrared Galaxy F00183-7111}

\author[Mao et al.]{Minnie~Y. Mao$^{1}$\thanks{e-mail: mmao@nrao.edu}, Ray P. Norris$^{2}$, Bjorn Emonts$^{2,3}$, Rob G. Sharp$^{4}$, 
\newauthor Ilana Feain$^{5}$, Kate Chow$^{2}$, Emil Lenc$^{5,6}$, Jamie Stevens$^{2}$\\
$^{1}$National Radio Astronomy Observatory, PO Box O, Socorro, NM, 87801, USA\\
$^{2}$CSIRO Astronomy and Space Science, PO Box 76, Epping, NSW, 1710, Australia\\
$^{3}$Centro de Astrobiolog\'{i}a (INTA-CSIC), Ctra de Torrej\'{o}n a Ajalvir, km 4, 28850, Torrej\'{o}n de Ardoz, Madrid, Spain\\
$^{4}$Research School of Astronomy and Astrophysics, The Australian National University, Cotter Road, Weston Creek,\\ACT, 2611, Australia\\
$^{5}$Sydney Institute for Astronomy, School of Physics, The University of Sydney, NSW 2006, Australia\\
$^{6}$ARC Centre of Excellence for All-sky Astrophysics (CAASTRO), NSW 2016, Australia\\
}
\begin{document}

\date{2013}

\pagerange{\pageref{firstpage}--\pageref{lastpage}} \pubyear{2013}

\maketitle

\label{firstpage}

\begin{abstract}
We report the detection of molecular CO(1-0) gas in F00183-7111, one of the most extreme Ultra-Luminous Infrared Galaxies known, with the Australia Telescope Compact Array.  We measure a redshift of 0.3292 for F00183-7111 from the CO(1-0) line and estimate the mass of the molecular gas in 00183 to be 1 $\times$ 10$^{10}$ M$_{\odot}$. We find that F00183-7111 is predominately powered by the AGN and only $\sim$14\,per cent of the total luminosity is contributed by star-formation (SFR $\sim$220\,M$_{\odot}$\,yr$^{-1}$). 

We also present an optical image of F00183-7111, which shows an extension to the East. We searched for star-formation in this extension using radio continuum observations but do not detect any. This suggests that the star-formation is likely to be predominately nuclear.

These observations provide additional support for a model in which the radio emission from ULIRGs is powered by an intense burst of star-formation and by a radio-loud AGN embedded in its nucleus, both triggered by a merger of gas-rich galaxies. 

\end{abstract}


\section{Introduction}
\label{intro}

Ultraluminous Infrared galaxies (ULIRGs) are a class of galaxy with a bolometric luminosity $>10^{12} {\rm L}_{\odot}$ \citep{Aaronson84, Houck85, Allen85}. They have been attributed to the merger of two gas-rich spirals \citep{Armus87,Sanders88, Veilleux02, Spoon09}, which fuels a pre-existing quiescent black hole and also triggers a powerful nuclear starburst. This intense starburst activity generates strong starburst-driven winds \citep{Rupke05, Heckman00}, which will eventually blow away the enshrouding dust and lay bare the quasar core, depleting the dust and gas to form an elliptical galaxy \citep{Dasyra06a, Dasyra06b}. Leading models suggest that the supermassive black hole grows by accretion while surrounded by a cocoon of dust \citep[e.g.][]{DiMatteo05, Hopkins05}, which is then shed by outflows driven by powerful quasar winds \citep{Balsara93}. This activity ceases when the fuel supply to the central regions is exhausted, and most of the remaining gas is expelled, starving both the active galactic nucleus (AGN) and the star-forming activity.  



In the low-redshift Universe, star-formation is dominated by M82-type starburst galaxies, and fewer than 50 ULIRGs are known at $z<0.1$. At higher redshifts, ULIRGs are much more common, and dominate the cosmic star-formation rate at $z \sim 2$  \citep[e.g][]{Casey12, Magnelli13}. ULIRGs suffer from many magnitudes of extinction to their nuclei, making it difficult to determine whether their dominant power source is due to an AGN or star-formation activity. Most ULIRGs in the local Universe appear to be  powered primarily by a starburst \citep[e.g.][]{Risaliti10}.


IRAS F00183-7111 (also known as IRAS 00182-7112, hereafter 00183) is one of the most luminous ULIRGs known, and lies at a redshift of 0.3276 \citep{Roy97}. Its bolometric luminosity is $9\times10^{12}$\,L$_\odot$ \citep{Spoon09}, most of which is radiated at far-infrared wavelengths. Previous infrared observations \citep{Rigopoulou99} and optical observations \citep{Drake04} show a disturbed morphology.

00183 contains a strong radio source \citep[108\,mJy at 4.8\,GHz:][]{Roy97} with a radio luminosity of L$_{4.8\,{\rm GHz}}=3\times10^{25}$W\,Hz$^{-1}$, which places it within the regime of high luminosity (FRII-class) radio galaxies. \citet{Norris12} have  obtained VLBI (Very Long Baseline Interferometry) data using the Long Baseline Array (LBA), which shows that 00183 contains a compact radio-loud AGN. The AGN accounts for nearly all of the radio luminosity of 00183, and has compact jets only 1.7\,kpc long. The morphology and spectral index are both consistent with Compact Steep Spectrum sources \citep[CSS,][]{Odea98}, which are widely thought to represent an early stage of evolution of radio galaxies \citep{Randall11}. \citet{Norris12} argue that these jets are boring through the dense gas and starburst activity that confine them.

This AGN is invisible at optical and near-infrared wavelengths because of the dense dust surrounding it, evidence for which includes  the deep 9.7\,$\mu$m silicate absorption feature \citep{Tran01, Spoon04}. However, the AGN is confirmed by the detection of a 6.7\,keV FeK line (Fe XXV) with a large equivalent width, indicative of reflected light from a Compton thick AGN \citep{Nandra07}.

00183 is believed to have been caught in the brief transition period between merging starburst and radio-loud `quasar-mode' accretion \citep[e.g.][]{Norris12}. The time-scales of the proposed ULIRG formation sequence are not well understood \citep[e.g.][]{Shabala09}, thus it is necessary to measure the relative contributions by star-formation and AGN to the total power of the ULIRG.

Observations of molecular gas can be used to show how the AGN is interacting with its host star-forming galaxy \citep[e.g.][]{Emonts11b, Rupke13}. Molecular hydrogen (H$_2$) is a key ingredient to forming stars, but unless shocked or heated to very high temperatures, H$_2$ is very difficult to detect due to its strongly forbidden rotational transitions. Fortunately, H$_2$ may be traced by carbon monoxide (CO), which emits strong rotational transition lines that occur primarily through collisions with H$_2$. CO traces the star-formation and is not contaminated by the presence of AGN. 

 In this Letter we present CO(1-0) observations, new optical observations and 6 and 9.5\,GHz radio continuum observations of 00183 in order to study the star-formation within this galaxy. Section 2 describes the observations and data analysis, Section 3 discusses the implications of the CO(1-0) detection and discusses star-formation in 00183 and Section 4 summarises our conclusions. 

Throughout this Letter, we assume H$_0$ = 71\,km\,s$^{-1}$\,Mpc$^{-1}$, $\Omega$$_M$ = 0.27 and $\Omega$$_\Lambda$ = 0.73 and we use the web-based calculator of \citet{Wright06} to estimate the physical parameters.

\section{Observations}\label{data}
\subsection{CO(1-0)}
Radio observations were carried out with the Australia Telescope Compact Array (ATCA, Project ID C2580) to search for CO(1-0) in 00183. The Compact Array Broadband Backend \citep[CABB,][]{Wilson11} was used in its coarsest spectral resolution mode with 2 $\times$ 2\,GHz of total bandwidth centred at 85.3 and 86.8\,GHz, and 1\,MHz spectral resolution in two linear polarizations. This corresponds to a velocity resolution of $\sim$3.5\,km\,s$^{-1}$ and an effective velocity coverage of $\sim$7000\,km\,s$^{-1}$. At $z\,=\,0.3276$ \citep{Roy97}, the optically measured redshift, the redshifted CO(1-0)\footnote{$\nu_{rest}$ = 115.2712\,GHz} should be detected at 86.827\,GHz. 

00183 was observed from 2011 October 02 -- 04 in the most compact hybrid H75 array configuration. This resulted in an observing time of 36 hours obtained under good weather, including calibration overheads of $\sim$40\,per cent.

Initial ATCA calibration, including bandpass calibration, was performed using 1921-293 and two minute scans of the phase calibrator J0103-6439\footnote{S$_{86.8\,{\rm GHz}}$ $\sim$0.59\,Jy\,beam$^{-1}$} were performed every ten minutes. Flux calibration was performed on Uranus, which was observed at a similar elevation to J0103-6439, in order to ensure we calibrated with the same air mass. 

The data were calibrated and imaged using \textsc{miriad} based on the method described in \citet{Emonts11a}. The spatial resolution of the final image is 7.2$\times$5.8 arcsec (P.A. = 85.9$^{\circ}$). For the data analysis presented here we have binned the data to 8\,MHz wide channels, which corresponds to a velocity resolution of 27.6\,km\,s$^{-1}$.

\subsection{Optical Image}

A snapshot image of 00183 was obtained with the Focal Plane Imager of the 2dF instrument at the Anglo-Australia Telescope during twilight on 2011 August 23. The seeing was 2.1\,arcsec. Four 2 minute dithered observations were combined after registration. Three 2 minute dark frames were used to dark-correct the image as the camera was operating at only 0$^{\circ}$C. The image is taken in a red filter which, when combined with the CCD sensitivity, corresponds broadly to an R+I filter and encompasses the wavelength of rest frame H-alpha emission from 00183.

\subsection{6 \& 9GHz}

Radio continuum observations of 00183 at 6 and 9.5\,GHz were carried out with the ATCA (Project ID C2749). The ATCA was used in standard continuum mode with 2 $\times$ 2\,GHz of total bandwidth centred at 6 and 9.5\,GHz in two linear polarizations. 

00183 was observed from 2013 February 24 -- 26, in the 6A configuration for a total of 36\,hours. This provided $\sim$1 arcsec spatial resolution over a field of view of $\sim$5 arcmin.

\section{Results and Discussion}
\begin{figure}
\begin{center}
\includegraphics[angle=0, scale=0.5]{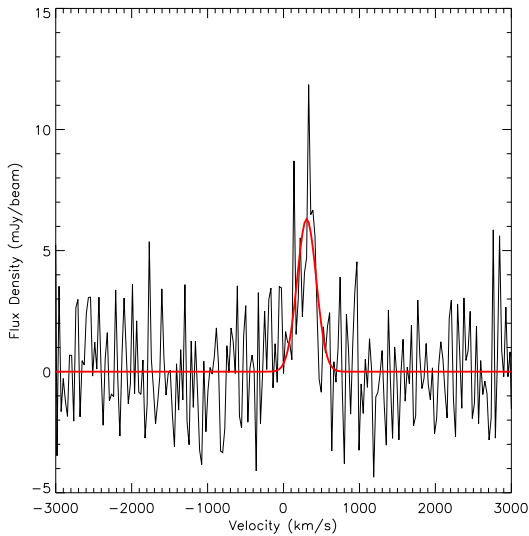}
\end{center}
\caption{CO(1-0) spectrum of 00183 after continuum subtraction. The spectrum has a velocity resolution of 27.6\,km\,s$^{-1}$. The red solid line shows the Gaussian fit to the line. } \label{spec}
\end{figure}

\begin{figure*}
\begin{center}
\begin{tabular}{ccc}
\includegraphics[angle=-90, scale=0.24]{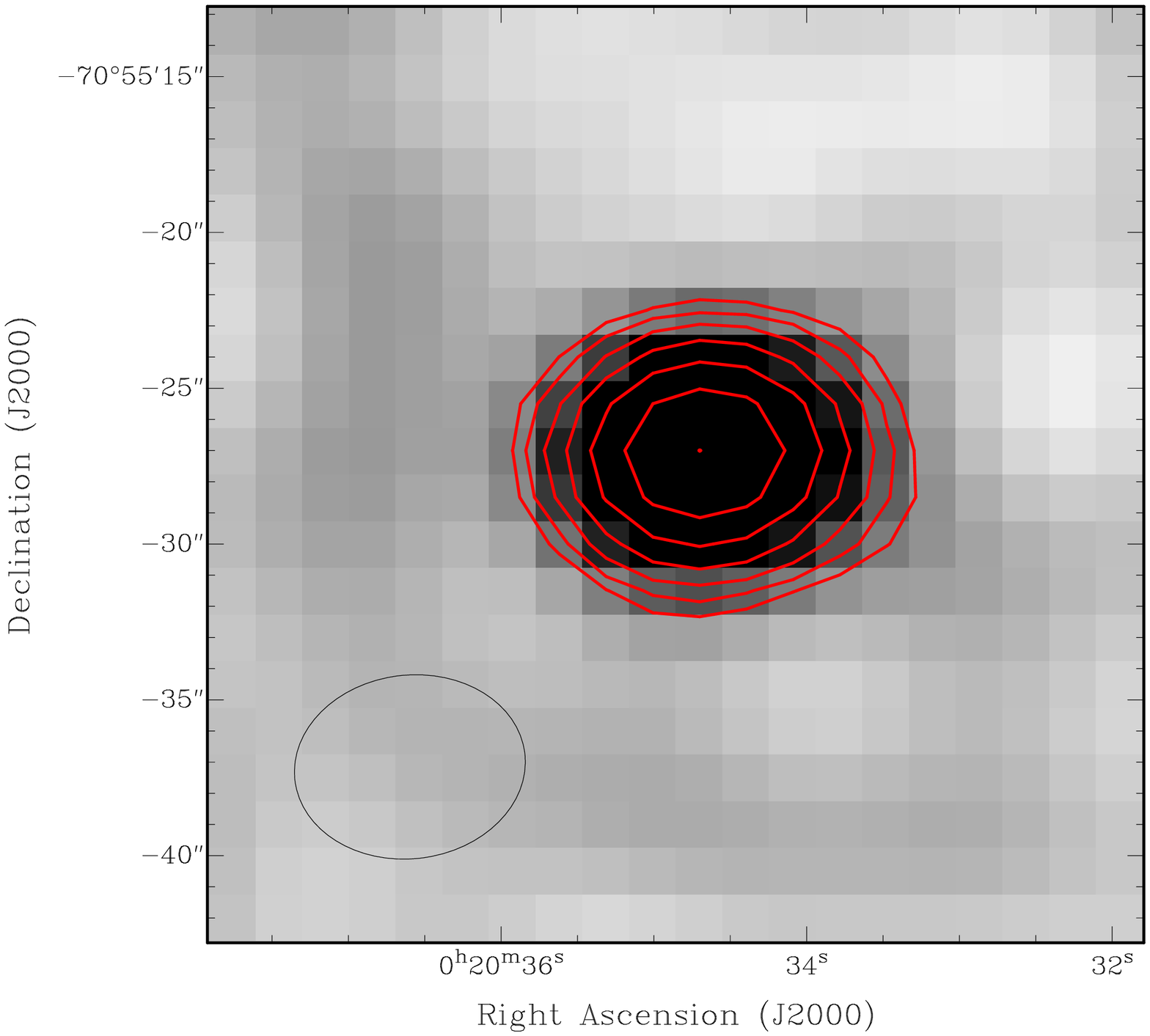}& \includegraphics[angle=-90, scale=0.24]{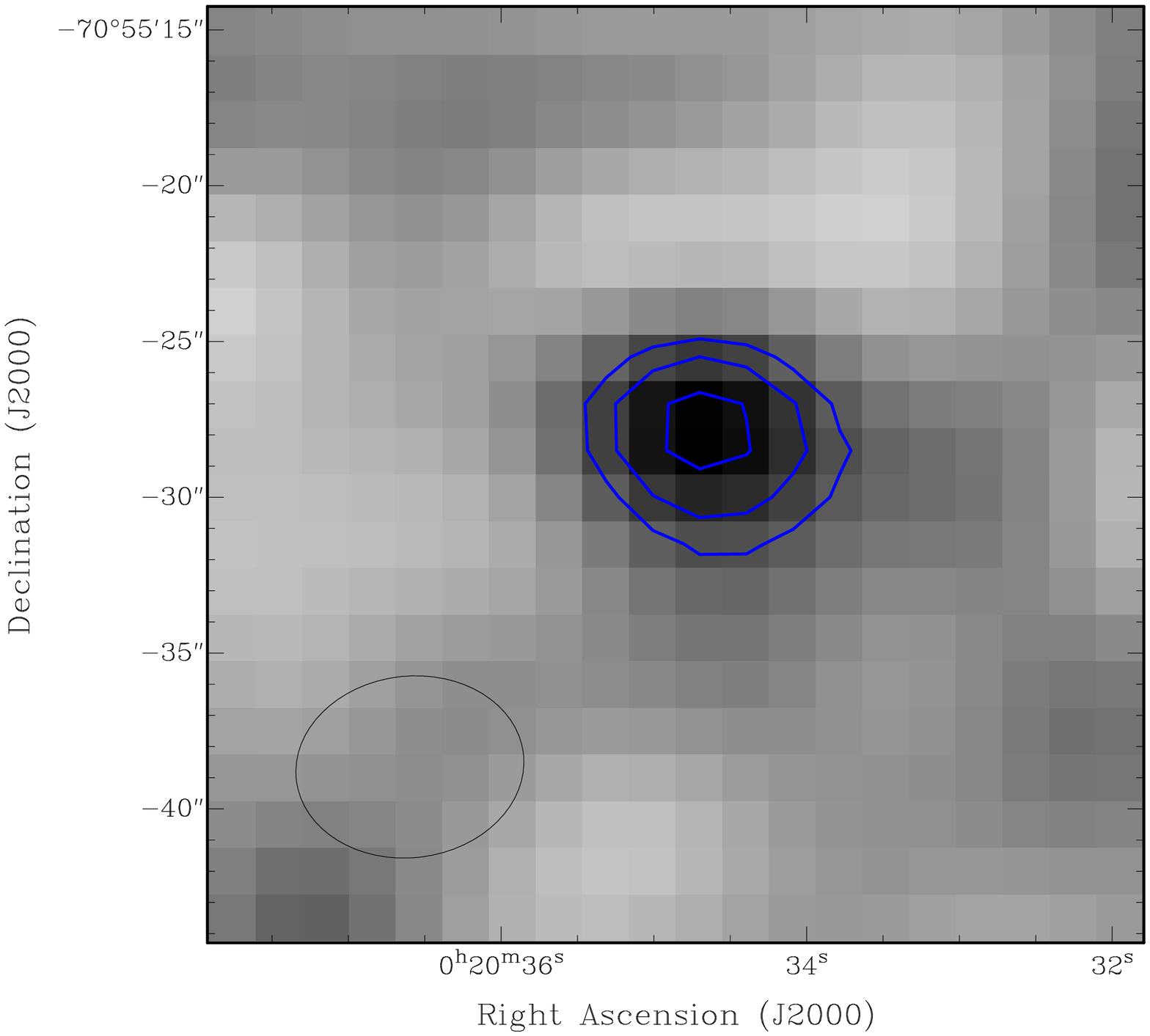}& \includegraphics[angle=-90, scale=0.24]{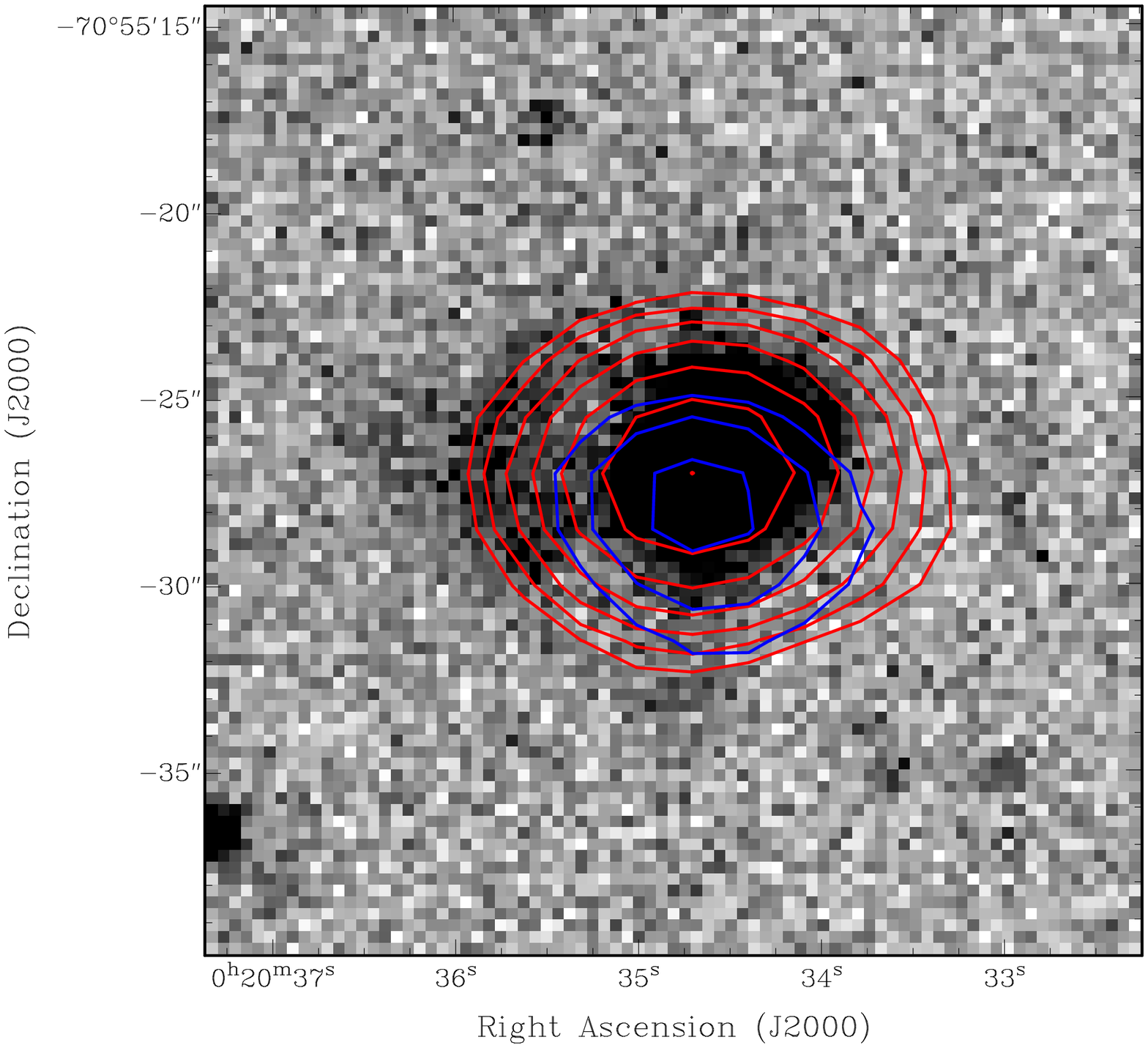} \\
\includegraphics[angle=-90, scale=0.24]{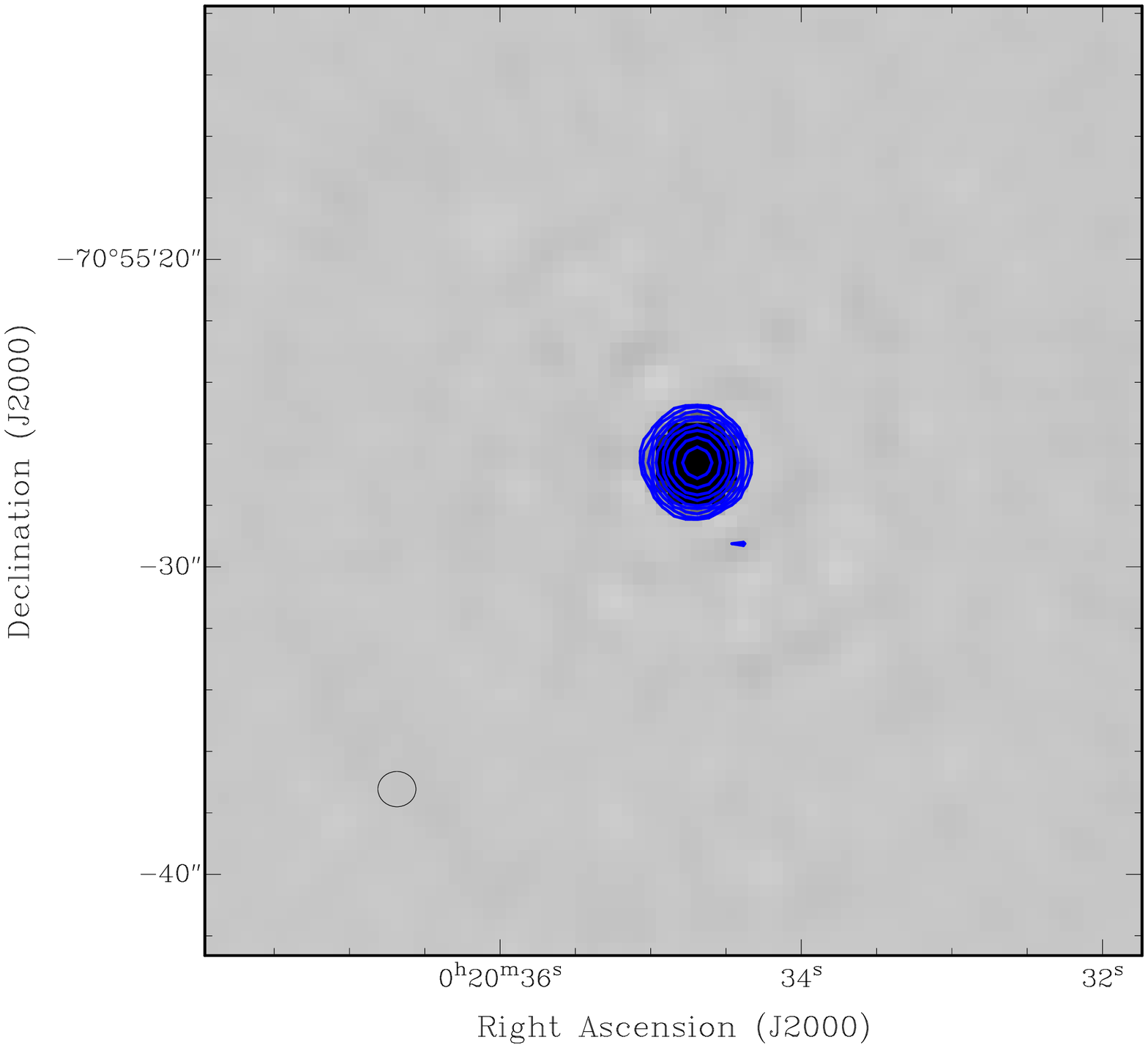}& \includegraphics[angle=-90, scale=0.24]{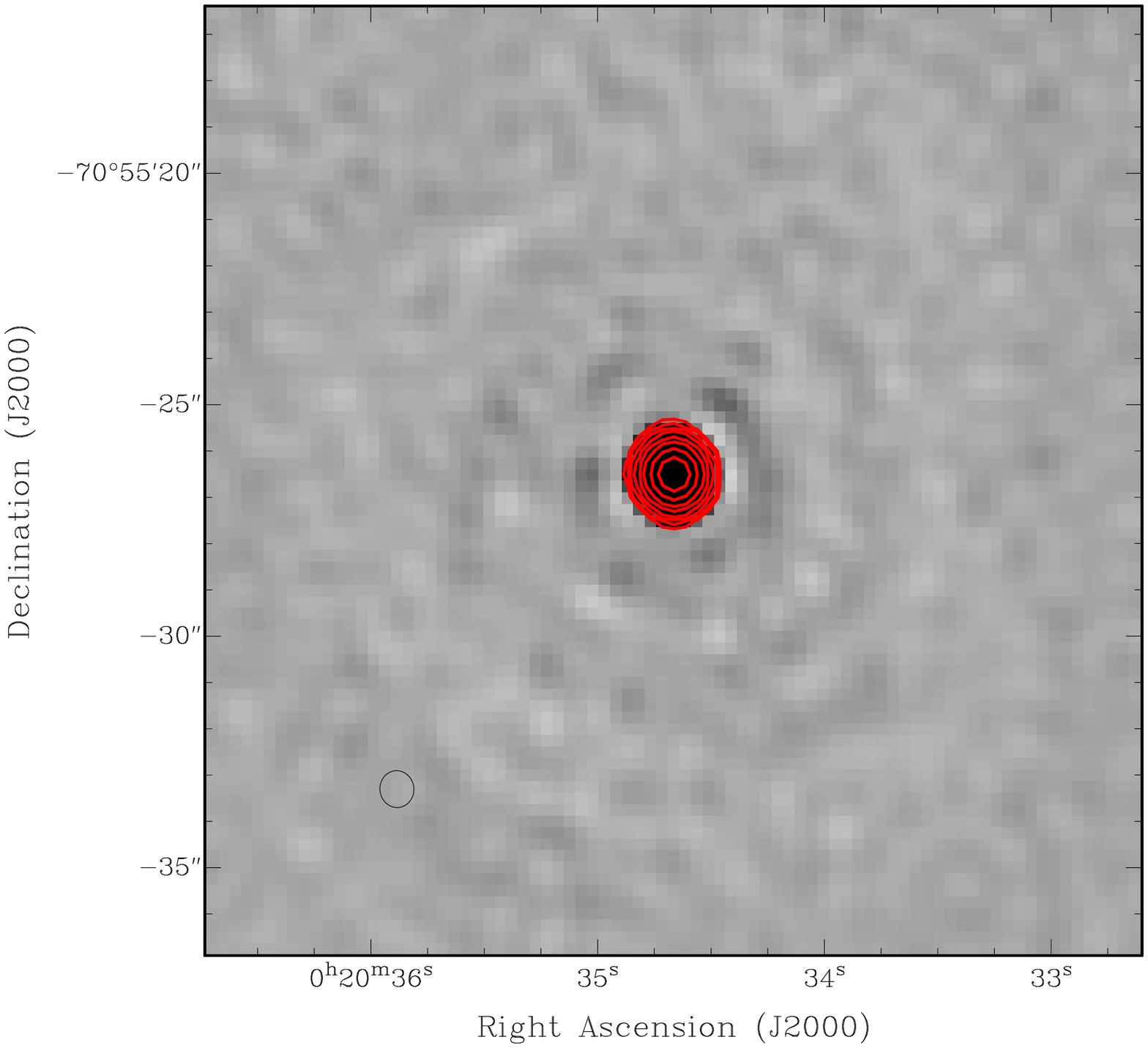}& \includegraphics[angle=-90, scale=0.24]{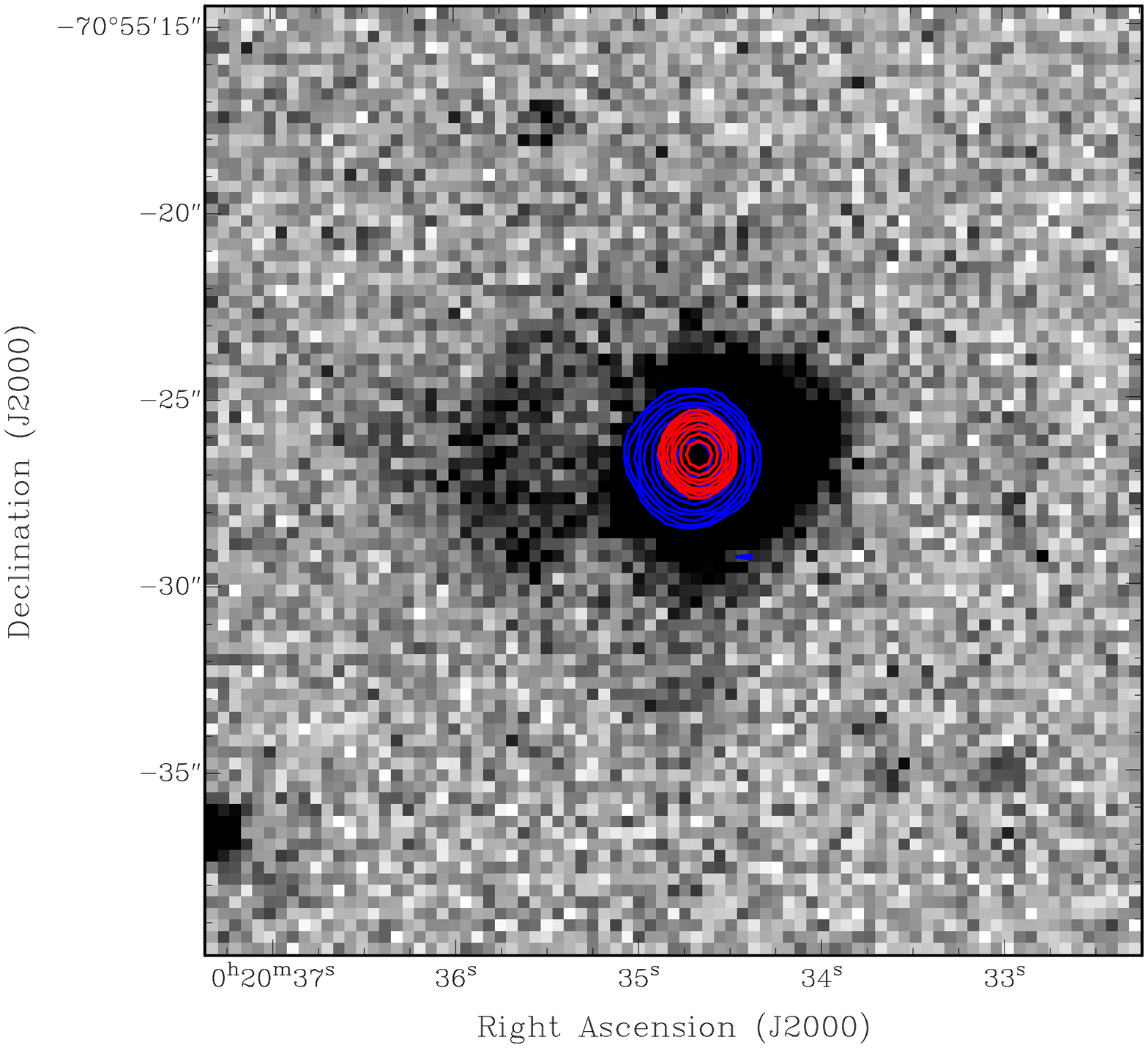} \\
\end{tabular}
\end{center}
\caption{\textbf{Top Left:} 3\,mm radio continuum greyscale image of 00183 overlaid with red contours. The contours start at 0.6\,mJy\,beam$^{-1}$ (3$\times$ rms) and increase by $\sqrt{2}$ intervals. The beam is shown in the lower left-hand corner. \textbf{Top Middle:} Total intensity (moment zero) image of the CO(1-0) detection shown in greyscale overlaid with blue contours. The contours start at 1\,Jy\,beam$^{-1}$km\,s$^{-1}$ (3$\times$ rms) and increase by $\sqrt{2}$ intervals. \textbf{Top Right:} Optical image shown in greyscale overlaid with the 3mm radio continuum contours (red) and the moment zero image contours (in blue). The contours are set at the same levels as the previous two panels. \textbf{Bottom Left:} 6\,GHz radio continuum greyscale image of 00183 overlaid with blue contours. The contours start at 0.2\,mJy\,beam$^{-1}$ (10$\times$ rms) and increase by $\times$2 intervals. \textbf{Bottom Middle:} 9.5\,GHz radio continuum greyscale image overlaid with red contours. The contours start at 0.2\,mJy\,beam$^{-1}$ (10$\times$ rms) and increase by $\times$2 intervals. \textbf{Bottom Right:} Optical image shown in greyscale overlaid with the 6\,GHz and 9.5\,GHz radio continuum contours. The contour colours and levels are the same as the previous two panels. }\label{image}
\end{figure*}

\subsection{CO(1-0) in 00183}

We obtain an unambiguous (7$\sigma$) detection of CO(1-0) in 00183 on top of radio continuum emission (Fig. \ref{spec}). Both the continuum and line emission are spatially unresolved (Fig. \ref{image}, top panels) and the CO(1-0) line detection is coincident with the continuum emission within $\sim$ 1 arcsec. After subtracting S$_{86.8\,{\rm GHz}}$ = 2.97\,mJy\,beam$^{-1}$ of continuum flux density from the spectrum, the CO(1-0) is clearly detected in 00183 at a central frequency of 86.72\,GHz, which corresponds to a redshift of 0.3292 (Fig. \ref{spec}). This implies a velocity difference of +305\,km\,s$^{-1}$ from the velocity derived from the optical spectrum \citep{Roy97}. We note that neither of these is necessarily the systemic velocity, and indeed the systemic velocity in such a chaotic system may be poorly defined. The CO velocity measured here traces the centre of mass of the molecular material. We fit a Gaussian to the spectrum and find that the CO(1-0) detection has a peak flux density of S$_{\rm CO}$ = 6.3\,mJy\,beam$^{-1}$ and FWHM = 297\,km\,s$^{-1}$.  

The integrated CO(1-0) signal is detected at a $\sim$7$\sigma$ level with $\int_{\rm v} {\rm S}_{\rm CO} \delta {\rm v}$ = 2.32 Jy\,beam$^{-1}$km\,s$^{-1}$ (Fig. \ref{image}, top middle panel). Using \citet{Solomon05}, the CO(1-0) luminosity may be calculated using:

\begin{equation}
{\rm L}'_{\rm CO} = 3.25 \times 10^7 (\frac{\int_{\rm v} {\rm S}_{\rm CO}\delta {\rm v}}{\rm Jy\,km\,s^{-1}}) (\frac{{\rm D}_{\rm L}}{\rm Mpc})^2 (\frac{\nu_{\rm rest}}{\rm GHz})^{-2} (1+z)^{-1},
\end{equation}

where L$'_{\rm CO}$ is expressed in K\,km\,s$^{-1}$\,pc$^2$. We calculate L$'_{\rm CO}$ for 00183 to be 1.25 $\times$ 10$^{10}$\,K\,km\,s$^{-1}$\,pc$^2$. 

We can estimate the mass of the molecular gas in 00183 from the CO(1-0) luminosity by applying the standard conversion factor $\alpha_{\rm X} = {\rm M}_{\rm H_2}/{\rm L}'_{\rm CO} = 0.8 {\rm M}_{\odot}$ (K\,km\,s$^{-1}$\,pc$^2$)$^{-1}$ for ULIRGs \citep{Downes98}. We note that the conversion of CO luminosity into molecular gas mass is dependent on the molecular gas conditions, such as its density, temperature and kinetic state \citep[e.g.][]{Glover11, Bolatto13, Mashian13}, and we acknowledge that this yields a conservative estimate for the molecular gas mass \citep[e.g.][]{Papadopoulos13}. Using this conversion factor, we estimate the mass of the molecular gas in 00183 to be 1 $\times$ 10$^{10}$ M$_{\odot}$.

We can determine the star-formation rate (SFR) of 00183 using the empirically determined relation between CO luminosity and IR luminosity from the starburst component \citep[e.g.][]{Carilli13}:

\begin{equation}
{\rm log}({\rm L}_{{\rm IR}_{\rm starburst}}) = 1.37 \times {\rm log}({\rm L}'_{\rm CO}) - 1.74,
\end{equation}

and 

\begin{equation}\label{SFR}
{\rm SFR} \sim \delta_{\rm MF} \times 1.0 \times 10^{-10}{\rm L}_{{\rm IR}_{\rm starburst}},
\end{equation}

where $\delta_{\rm MF}$ = 1.8 was used assuming a Salpeter IMF \citep[e.g][]{Kennicutt98}. From these we calculate a SFR $\sim$220\,M$_{\odot}$\,yr$^{-1}$ for 00183. 

The bolometric luminosity of 00183 is  L$_{8 - 1000\mu {\rm m}}$ = $9 \times10^{12}$ L$_\odot$ \citep{Spoon09}, most of which is radiated at far-infrared wavelengths. Using Equation \ref{SFR}, this would imply a SFR of $\sim$1600\,M$_{\odot}$\,yr$^{-1}$. Moreover, 00183 has a hard X-ray luminosity of L$'_{2 - 10\,{\rm keV}}$ = $2 \times10^{44}$ ergs$^{-1}$ \citep{Nandra07}, which, if assuming 00183 is solely powered by star-formation, would imply a SFR of $>$12000\,M$_{\odot}$\,yr$^{-1}$. 

We know that 00183 harbours a powerful AGN in its core. \citet{Norris12} have detected and resolved the AGN, which accounts for nearly all of the radio luminosity of 00183. Furthermore, \citet{Nandra07} calculate that the AGN accounts for $>$80\,per cent of the total IR luminosity. This is in agreement with previous work by \citet{Spoon04}, who find that star-formation contributes less than 30\,per cent of the total IR luminosity. \citet{Ranalli03} had previously inferred that 00183 has a SFR of $\sim$310\,M$_{\odot}$\,yr$^{-1}$ from soft X-ray data. Consequently our derived SFR of $\sim$220\,M$_{\odot}$\,yr$^{-1}$ from the CO(1-0) appears consistent with previous work. This suggests that only $\sim$14\,per cent of the total power of 00183 is contributed by star-formation. This is particularly interesting as most ULIRGs appear to be powered predominately by star-formation \citep[e.g][]{Genzel98,Armus07}. \citet{Veilleux09} find that the average AGN contribution to the bolometric luminosity of ULIRGs is $\sim$35--40\,per cent, and they observe a trend with AGN contribution and merger-stage. It is likely that 00183 is in a late-merger phase where gas and dust has been blown out, decreasing active star-formation, consistent with the scenario proposed by \citet{Norris12}.

Both IR and X-ray star-formation diagnostics may be `contaminated' by AGN. CO is generally a good indicator of star-formation and is not contaminated by AGN. This is particularly true for the ground transition CO(1-0), as it is least affected by potential excitation conditions in the nuclear AGN region. The detection of CO(1-0) has provided the `cleanest' diagnostic yet for the star-formation in 00183.

\subsection{Radio Continuum from Star-formation}

A deep optical image of 00183 was obtained using the AAT (Fig. \ref{image}, right-most panels). There appears to be a $\sim$5\,arcsec extension to the East of the galaxy. 

The CO detection is located in the centre of 00183 and appears unresolved, thus it is likely that most of the star-formation is nuclear (within $\sim$5\,kpc). Nonetheless, we embarked upon a project to determine if there was any strong star-formation in the optical extension. If we assume that $\sim$14\,per cent of the IR luminosity is due to star-formation, then using the far-infrared radio correlation we can estimate the contribution by star-formation to the radio continuum emission. The 70\,$\mu$m flux density is S$_{70\mu {\rm m}}$ = 1.5\,Jy, 14\,per cent of which is $\sim$0.21\,Jy. Using q$_{70}$ = 2.15 \citep{Appleton04}, we would then expect to detect 1.5\,mJy at 1.4\,GHz of radio continuum that is due to star-formation. Assuming $\alpha^{8.6\,{\rm GHz}}_{1.4\,{\rm GHz}}$=-0.97 \citep{Drake04}\footnote{Spectral index is defined as S $\propto$ $\nu^{\alpha}$.}, this corresponds to S$_{6\,{\rm GHz}} \sim 370\,\mu$Jy and S$_{9.5\,{\rm GHz}} \sim 230\,\mu$Jy.

Fig. \ref{image} (bottom panels) shows the radio continuum images of 00183 at 6\,GHz and 9.5\,GHz. As these data were obtained in a single telescope configuration, 6A, there are some gaps in the \emph{uv}-plane, which have caused imaging artefacts to be present in the immediate vicinity of the bright source. As such, while the rms at both frequencies is $\sim$6\,$\mu$Jy\,beam$^{-1}$ throughout most of the image, the rms increases to $\sim$20\,$\mu$Jy\,beam$^{-1}$ next to 00183. At both these frequencies 00183 appears unresolved down to an rms of $\sim$20\,$\mu$Jy\,beam$^{-1}$ with S$_{6\,{\rm GHz}}$ = 81.7\,mJy\,beam$^{-1}$ and S$_{9.5\,{\rm GHz}}$ = 45.6\,mJy\,beam$^{-1}$. This corresponds to a $\alpha^{9.5\,{\rm GHz}}_{6\,{\rm GHz}} \sim -1.27$, which is consistent with $\alpha^{8.6\,{\rm GHz}}_{4.8\,{\rm GHz}}=-1.27$ measured by \citet{Drake04}. 

We measure an upper limit of S$_{6\,{\rm GHz}} \sim 100\,\mu$Jy to the radio continuum at the location of the optical extension and calculate an upper limit to the SFR in this region as $<$60\,M$_{\odot}$\,yr$^{-1}$. The unresolved radio detection suggests that the star-formation is likely to be nuclear and the lack of extended radio emission rules out the possibility of vast amounts of star-formation occurring outside of the nucleus.

\section{ Conclusions}
00183, one of the most luminous ULIRGs known, is believed to have been caught in the brief transition period between merging starburst and `quasar-mode' accretion. We have detected CO(1-0) in 00183 implying a SFR of 220\,M$_{\odot}$\,yr$^{-1}$, which is consistent with previous estimates based on infrared and X-ray diagnostics. This suggests that only $\sim$14\,per cent of the total power of the source is contributed by star-formation. 

The powerful AGN that dominates the power of 00183 is believed to be a very young quasar that has only just turned on and is in the process of quenching and turning off the star-formation in this ULIRG. These results provide additional support for a model in which ULIRGs are powered by star-formation and AGN activity, both of which are triggered by a merger of gas-rich galaxies. 

\section*{Acknowledgements}
We thank the staff at the ATCA and the AAO (especially Steve Lee!) for making these observations possible. We also thank our Referee, Joan Wrobel, Chris Carilli and Juergen Ott for their helpful comments. The ATCA is part of the ATNF, which is funded by the Commonwealth of Australia for operation as a National Facility managed by CSIRO. This research has also made use of NASA's Astrophysics Data System. The Centre for All-sky Astrophysics (CAASTRO) is an Australian Research Council Centre of Excellence, funded by grant CE110001020.

\clearpage

\label{lastpage}
\end{document}